\documentclass[journal]{IEEEtran}
\usepackage{amsmath,amsfonts}
\usepackage{amsthm}
\usepackage{algorithmic}
\usepackage{algorithm}
\usepackage{array}
\usepackage[caption=false,font=normalsize,labelfont=sf,textfont=sf]{subfig}
\usepackage{textcomp}
\usepackage{stfloats}
\usepackage{url}
\usepackage{verbatim}
\usepackage{graphicx}
\usepackage{cite}
\usepackage{balance}
\usepackage{upgreek}
\usepackage{color}
\newtheorem{theorem}{Theorem}
\newtheorem{lemma}{Lemma}
\newtheorem{remark}{Remark}
\newtheorem{corollary}{Corollary}

\begin{document}
	
\title{Achieving Full Multipath Diversity by Random Constellation Rotation: a Theoretical Perspective}
	
\author{
	Xuehan~Wang,~\IEEEmembership{Graduate~Student~Member,~IEEE},
	Jinhong~Yuan,~\IEEEmembership{Fellow,~IEEE}, 
	Jintao~Wang,~\IEEEmembership{Fellow,~IEEE}, and
	Kehan~Huang,~\IEEEmembership{Graduate~Student~Member,~IEEE}	
	\thanks{
		This work was supported in part by ZTE Industry-University-Institute Cooperation Funds under Grant No. IA2025012012. \textit{(Corresponding author: Jintao Wang.)}
		\par 
		Xuehan Wang and Jintao Wang are with the Department of Electronic Engineering, Tsinghua University, Beijing 100084, China, and also with the State Key laboratory of Space Network and Communications, Tsinghua University, Beijing 100084, China (e-mail: wang-xh21@mails.tsinghua.edu.cn; wangjintao@tsinghua.edu.cn).\par 
		Jinhong Yuan and Kehan Huang are with the School of Electrical Engineering and Telecommunications, University of New South Wales, Sydney, NSW 2052, Australia (e-mail: j.yuan@unsw.edu.au; kehan.huang@unsw.edu.au).\par 
	}
}
	
	%
	
	\maketitle
	
\begin{abstract}
	Diversity is an essential concept associated with communication reliability in multipath channels since it determines the slope of bit error rate performance in the medium to high signal-to-noise ratio regions. However, most of the existing analytical frameworks were developed for specific modulation schemes while the efficient validation of full multipath diversity for general modulation schemes remains an open problem. To fill this research gap, we propose to utilize random constellation rotation to ease the conditions for full-diversity modulation designs. For linearly precoded cyclic-prefix orthogonal frequency division multiplexing (OFDM) systems, we prove that maximum multipath diversity can be attained as long as the spread matrix does not have zero entries, which is a sufficient but easily satisfied condition. Furthermore, we derive the sufficient and necessary condition for general modulation schemes, whose verification can be divided into validation tasks for each column of the modulation matrix. Based on the proposed conditions, maximum diversity order can be attained with the probability of 1 by enabling a randomly generated rotation pattern for both time and doubly dispersive channels. The theoretical analysis in this paper also demonstrates that the diversity evaluation can be concentrated on the pairwise error probability when the number of error symbols is one, which reduces the complexity of diversity-driven design and performance analysis for novel modulation schemes significantly in both time and doubly dispersive channels. Finally, numerical results for various modulation schemes confirm that the theoretical analysis holds in both time and doubly dispersive channels. Furthermore, when employing practical detectors, the random constellation rotation technique consistently enhance the transmission reliability for both coded and uncoded systems.
\end{abstract}

\begin{IEEEkeywords}
	Diversity, multipath channels, maximum likelihood (ML), random constellation rotation
\end{IEEEkeywords}
	
\section{Introduction}
\label{sec_intro}
Next-generation wireless networks are expected to realize ultra-reliable communications in both time and doubly dispersive channels \cite{NGcommunication_multipath,OTFS_STSK_diversity,TSP_multipath_OFDM,TSP_multipath_VOFDM}. This indicates that modulation schemes are required to achieve full multipath diversity since it offers valuable countermeasures against fading \cite{TFdouble_Vandermonde,diversity_multiplex_tradeoff_ISI} by revealing the maximum slope of bit error rate (BER) performance in the medium to high signal-to-noise ratio (SNR) regions. As a result, it is essential to establish an efficient method to determine the diversity order achieved by general modulation schemes and develop novel transmission approaches accordingly.\par
To achieve this target, many modulation strategies have been investigated in the literature. For time dispersive channels, the authors in \cite{FrequencyNonzero_first_SPAWC,FrequencyNonzero_TIT} first provided the sufficient and necessary conditions to achieve full diversity for linearly precoded cyclic prefix orthogonal frequency division multiplexing (CP-OFDM) systems. The discrete Fourier transform spread OFDM (DFT-s-OFDM) with constellation rotation obtained by computer searching was then developed in \cite{phase_rota_DFTsOFDM}. An elaborately designed Vandermonde matrix can also serve as a qualified precoding matrix as indicated in \cite{VandermondeLCP_OFDM_delayOnly}. When considering the doubly dispersive channels, the general criterion for maximum diversity order can be simplified by adding enough zero-padding samples \cite{TFdouble_Vandermonde,ZP_fullDiversity}. On the other hand, some emerging candidates also reveal the optimal diversity order under specific parameter settings. For example, orthogonal time frequency space (OTFS) and orthogonal delay-Doppler division multiplexing (ODDM) systems approach maximum diversity with a specified phase rotation pattern \cite{ref_ODDM,OTFS_diversity_TWC,OTFS_diversity_WCL,shuangyang_OTFS} while affine frequency division multiplexing (AFDM) schemes can also enjoy full diversity order if both chirp parameters are set appropriately \cite{ref_AFDM,AFDM_diversity_JSAC}.\par 
However, most of the existing analysis is dedicated to specific modulation schemes and the approaches utilized are typically not applicable for other systems. To investigate whether full diversity order has been achieved for a general modulation system, the complexity required grows exponentially with the number of data symbols if employing the definition in \cite{ref_LTVchannel_book} directly. As a result, the analytical framework for multipath diversity is still not sufficient for inspiring and evaluating novel modulation schemes for emerging applications. On the other hand, the separability has been demonstrated to be helpful for promoting system reliability \cite{ZP_fullDiversity,ref_AFDM,ref_ODDM,OTFS_diversity_LP_WCL}, i.e., the error performance can be significantly enhanced if the demodulated channel-distorted elements from different equivalent channel taps do not overlap for each data component. According to the pairwise error probability (PEP) framework for defining diversity, separability corresponds to attaining full diversity when exactly one data symbol is erroneous. Therefore, it is possible to establish the connection between the characteristics of the transmitted sequence for each data element and whether full diversity order can be expected, which can benefit the research community of modulation schemes significantly.\par 
To address this issue, we propose employing the random constellation rotation to achieve maximum multipath diversity in this paper inspired by the application of phase rotation in \cite{phase_rota_DFTsOFDM,OTFS_diversity_TWC,rotation_VOFDM}. We first develop the theoretical analysis for time dispersive channels and then present the extension to doubly dispersive channels. Both linearly precoded CP-OFDM and general modulation systems are investigated in this paper, where the conditions for optimal diversity order are provided with theoretical proof. Considering that the constellation rotation only introduces linear complexity, the overall computational load for verifying the optimality of diversity of general modulation schemes can also be reduced to a linear order, which is helpful for inspiring novel modulation designs and the corresponding performance analysis\footnote{The proposed approach is not a specific scheme for conventional multicarrier systems, e.g., OFDM. The analysis is also leveraged for general linear modulation systems epitomized by other state-of-the-art waveforms, e.g., OTFS and orthogonal time sequency multiplexing (OTSM) \cite{ref_OTSM}.}. The contributions of this paper can be summarized as follows:\par
\begin{itemize}
	\item For linearly precoded CP-OFDM systems, we prove in \textbf{Theorem \ref{th1_spread}} that as long as the spread matrix includes \textbf{no zero entries}, full multipath diversity can be achieved for time dispersive channels with infinite rotation choices. Though this is a sufficient but not necessary condition, it is easily satisfied in practical designs.
	\item For general linear modulation systems, we prove in \textbf{Theorem \ref{th2_modu}} that the sufficient and necessary condition for full diversity in time dispersive channels only relies on the property of \textbf{the transmitted sequence for each data symbol}, i.e., full rank for PEP analysis when \textbf{the number of error symbols is one} can already guarantee the maximum diversity order.
	\item For both linearly precoded CP-OFDM and general linear modulation systems satisfying the corresponding conditions in time dispersive channels, we have proposed an algorithm to construct rotation angles to achieve strictly full diversity. Besides, we also prove that if the rotation angles are generated \textbf{randomly}, optimal diversity order can be expected with \textbf{the probability of 1}. Meanwhile, the theoretical analysis for both schemes can be extended to doubly dispersive channels.
	\item Various application and extension of the analysis in this paper are presented. For example, we apply \textbf{Theorem \ref{th1_spread}} to demonstrate the superiority of DFT-s-OFDM systems with random constellation rotation over time dispersive channels. We also validate that if the last prefix element of the transmitted sequence can be adjusted, all modulation schemes can achieve the diversity order of \textbf{at least 2}. \textbf{Corollary \ref{corollary_doubly_general}} extends the analysis in \textbf{Theorem \ref{th2_modu}} for time dispersive channels to doubly dispersive channels, which explains the optimal error performance of several emerging modulation strategies such as OTFS, ODDM, and AFDM under doubly-dispersive channels by establishing the bridge between full diversity and separability.
\end{itemize}\par 
It is worth pointing out that the theoretical analysis in this paper reveals that the diversity analysis can be concentrated on the investigation of PEP with only one error symbol by enabling a random constellation rotation pattern independent of the specific channel and data realization. It contributes to the development of diversity theory significantly since the standard evaluation for a general modulation scheme requires traversing PEP for all possible error events, whose complexity grows exponentially with the number of data symbols within a modulation frame. It is also beneficial for understanding and inspiring novel linear modulation schemes since the designing and performance analysis can be simplified significantly for both time and doubly dispersive channels if full multipath diversity is admired. We also demonstrate that randomly generated rotation patterns enjoy the probability of 1 to achieve full diversity if the conditions on PEPs with only one error data symbol have been satisfied. Nevertheless, prior investigations can only design the rotation patterns for specific modulation schemes, e.g., the computer search-based results in \cite{phase_rota_DFTsOFDM} for DFT-s-OFDM systems. Moreover, this insight can be extended to doubly dispersive channels from both theoretical analysis and numerical results.\par 
The rest of this paper is organized as follows. The preliminaries about the communication system models in time and doubly dispersive channels, and multipath diversity for general linear modulation schemes are briefly reviewed in Section \ref{sec_model}. The theoretical derivations for linearly precoded CP-OFDM systems and general modulation schemes are presented in Sections \ref{sec_spread} and \ref{sec_modulation}, respectively. Numerical results are provided in Section \ref{sec_simu} to demonstrate the analysis in this paper. Finally, conclusions are briefly drawn in Section \ref{sec_conclusion}.\par 
\textit{Notations}: $\mathbf{A}$ is a matrix, $\mathbf{a}$ is a column vector, $a$ is a scalar, $\mathcal{A}$ is a set. $\mathbf{A}^{T}$ and $\mathbf{A}^{H}$ denote the transposition and conjugate transposition. $\mathbf{A}=\text{diag}\left(a_{0},a_{1},\cdots,a_{N-1}\right)$ denotes a diagonal matrix of size $N$ with $\left(a_{0},a_{1},\cdots,a_{N-1}\right)$ as its diagonal elements. $(\cdot)^{*}$ denotes the conjugate operation and $\mathbb{E}\left\{\cdot\right\}$ represents the expectation operation. $||\cdot||$ is the $l_{2}-$norm. $\mathbf{I}_{M}$ and $\mathbf{0}_{M\times{N}}$ denote the identity matrix of size $M$, $M\times{N}$ matrix of zeros, respectively. Finally, let $\mathcal{A}-\mathcal{A}=\left\{x-y|x\in\mathcal{A}, y\in\mathcal{A}\right\}$ represent the possible error scenarios for a mapping alphabet $\mathcal{A}$ to ease the PEP illustration.\par

\section{Preliminaries}
\label{sec_model}
In this section, the preliminaries for multipath diversity are presented to ease the illustration. A general communication system model in time and doubly dispersive channels is first introduced. Following the illustrations in \cite{TFdouble_Vandermonde,group_OFDM,FrequencyNonzero_TIT,OTFS_diversity_TWC,ref_AFDM,OTFS_diversity_WCL}, the modulation and demodulation matrices are employed to characterize the general linear modulation systems at the sequence level. Then the basic concepts of multipath diversity are briefly reviewed to establish the basis for the analysis in this paper.
\subsection{System Model}
\label{subsec_channel}
In this paper, the baseband equivalent model for time dispersive channels is considered with inherent sparsity as in \cite{ref_LTVchannel_book,ref_channel,ref_ODDM}, where the baseband continuous signal $s(t)$ is sent from the transmitter to the receiver via $I$ incident paths. As a result, the received continuous signal can be derived as
\begin{equation}
	r(t)=\sum_{i=1}^{I}\tilde{h}_{i}s(t-\tau_{i})+w(t),
	\label{LTI_continuous}
\end{equation}
where $\tilde{h}_{i}$ and $\tau_{i}$ denote the complex gain and delay associated with the $i$-th path, and $w(t)$ stands for the additive noise at time $t$, respectively. To ease the system design, an equivalent sampled model is usually considered by sampling $s(t)$ and $r(t)$ with the sampling period $T_{s}$ under the constraints of Nyquist criteria to get the time domain transmitted and received samples as $s[p]$ and $r[p]$. Let $L$ denote the maximum delay tap for the equivalent sampled channel, which includes the spread for both the multipath channel and pulse-shaping. The time domain input-output relation between discrete samples can then be derived \cite{ref_ODDM,ref_LTVchannel_book,ref_channel,FrequencyNonzero_TIT,phase_rota_DFTsOFDM} as
\begin{equation}
	r[p]=\sum_{l=0}^{L-1}h_{l}s[p-l]+w[p],
	\label{IO_delayOnly}
\end{equation}
where $h_{l}$ and $w[p]\sim\mathcal{CN}\left(0,N_{0}\right)$ represent the equivalent complex gain of the $l$-th delay tap and additive white Gaussian noise (AWGN) samples, respectively. $N_{0}$ denotes the noise energy per complex symbol.\par 
Let $\{s[p]|p=-M_{p},-M_{p}+1,\cdots,M-1\}$ denote the transmitted sequence within a modulation frame, where $M$ and $M_{p}\geq L$ represent the number of data symbols and the length of prefix\footnote{Similar to the proof in \cite{ref_LTVchannel_book,ZP_fullDiversity}, if the prefix is padded by zeros, arbitrary modulation schemes can achieve full diversity under time dispersive channels. For zero-padded (ZP)-OFDM systems, the orthogonality among subcarriers is disrupted in multipath channels. Therefore, the transmitted symbols on each subcarrier yield receiving components across multiple subcarriers, leading to diversity greater than 1. Therefore, we focus on the scenarios when the prefix includes nonzero values in this paper.} samples for avoiding the interference between adjacent modulation frames considering the input-output relation in \eqref{IO_delayOnly}. The vectorized input-output characterization between the transmitted and received sequence can then be derived as 
\begin{equation}
	\mathbf{r}=\sum_{l=0}^{L-1}h_{l}\mathbf{A}_{l}\mathbf{s}+\mathbf{w},
	\label{vector_IO}
\end{equation}
where the $p$-th element of $\mathbf{r},\mathbf{w}\in\mathbb{C}^{M\times1}$ are $r[p]$ and $w[p]$ for $p=0,1,\cdots,M-1$, respectively. $\mathbf{s}\in\mathbb{C}^{(M+M_{p})\times1}$ denotes the vectorized transmitted sequence whose $p$-th element is $s[p]$ for $p=-M_{p},-M_{p}+1,\cdots,M-1$. Finally, $\mathbf{A}_{l}$ represents the time-domain channel matrix associated with the $l$-th delay tap, which can be derived as $\mathbf{A}_{l}=\left[\mathbf{0}_{M\times(M_{p}-l)},\mathbf{I}_{M},\mathbf{0}_{M\times{l}}\right]$.\par 
The impact of modulation is then taken into consideration based on \eqref{vector_IO}. To achieve full multipath diversity, the techniques of constellation rotation are adopted like \cite{phase_rota_DFTsOFDM}. To be more specific, the data vector lies in $e^{j\phi_{0}}\mathcal{A}\times e^{j\phi_{1}}\mathcal{A}\times\cdots\times e^{j\phi_{M-1}}\mathcal{A}$, where $\mathcal{A}$ represents the mapping alphabet to load data bits while $\phi_{0},\phi_{1},\cdots,\phi_{M-1}$ are the rotation angles to be specified. Let $\mathbf{d}\in\mathcal{A}^{M}$ denote the initial data vector without phase rotation, the transmitted sequence can be derived by $\mathbf{s}=\mathbf{\Psi\Phi}\mathbf{d}$, where $\mathbf{\Psi}\in\mathbb{C}^{(M+M_{p})\times{M}}$ and $\mathbf{\Phi}=\text{diag}(e^{j\phi_{0}},e^{j\phi_{1}},\cdots,e^{j\phi_{M-1}})$ represent the modulation matrix employed for generating the transmitted sequence and phase rotation matrix, respectively. Let $\mathbf{G}\in\mathbb{C}^{M\times{M}}$ denote the corresponding demodulation matrix\footnote{For most scenarios, the modulation is carried out via two steps as orthogonal transform represented by the unitary matrix $\dot{\mathbf{\Psi}}$ followed by the prefix addition, where $\dot{\mathbf{\Psi}}\in\mathbb{C}^{M\times M}$ is the last $M$ rows of $\mathbf{\Psi}$. Then we can derive $\mathbf{G}=\dot{\mathbf{\Psi}}^{H}$. Please note that the prefix might not be a cyclic prefix, e.g., AFDM systems \cite{AFDM_diversity_JSAC,ref_AFDM}.}, the input-output relation can be equivalently rewritten by
\begin{equation}
	\mathbf{y}=\mathbf{Gr}=\sum_{l=0}^{L-1}h_{l}\mathbf{H}_{l}\mathbf{d}+\tilde{\mathbf{w}},
\end{equation}
where\footnote{For practical realization, we can also denote $\mathbf{\Psi}_{1}=\mathbf{\Psi\Phi}$ as the modulation matrix. However, the decomposition here provides significant convenience for diversity analysis and achievement, which will be depicted in detail in the following sections.} $\mathbf{H}_{l}=\mathbf{GA}_{l}\mathbf{\Psi\Phi}$ denotes the equivalent channel matrix associated with the $l$-th tap.\par 
On the other hand, general channel conditions will involve not only time dispersion but also frequency dispersion, which leads to $2K+1$ Doppler taps within each delay tap, where $K$ is the maximum sampled Doppler tap. As a result, the input-output relation in \eqref{IO_delayOnly} can be modified \cite{TFdouble_Vandermonde,ref_channel} as
\begin{equation}
	r[p]=\sum_{l=0}^{L-1}\sum_{k=-K}^{K}h_{l}^{k}e^{j2\pi\frac{pk}{M}}s[p-l]+w[p],
	\label{IO_double}
\end{equation}
where $h_{k}^{l}$ and $M$ denote the complex gain for the $k$-th Doppler tap within the $l$-th delay tap and the length\footnote{In this paper, we assume that the length of the received sequence is equal to the number of data symbols to simplify the illustration. It is trivial to extend the analysis to fewer data symbols for both linearly precoded CP-OFDM systems and general linear modulation schemes.} of the received sequence, respectively. In this paper, we mainly focus on the derivations for time dispersive channels as in \eqref{IO_delayOnly}. However, our following analysis over time dispersive channels can be extended to doubly dispersive channels with $(2K+1)L<M$ like \cite{TFdouble_Vandermonde}, which can be seen clearly in Section \ref{subsec_modulation_discussion}.
\subsection{Diversity}
\label{subsec_diversity}
In this subsection, an analytical framework for multipath diversity is provided as the basis. Without loss of generality, we assume $\mathbf{G}$ is a unitary matrix\footnote{The derivations can be easily extended to scenarios when $\mathbf{G}$ is non-singular.} to ensure $\tilde{\mathbf{w}}\sim\mathcal{CN}(\mathbf{0},N_{0}\mathbf{I}_{M})$ and $h_{l}\overset{\text{i.i.d.}}{\sim}\mathcal{CN}(0,\frac{1}{P})$. Let $\mathbf{d}_{1}\rightarrow\mathbf{d}_{2}$ define the pairwise error event with $\mathbf{d}_{1}\neq \mathbf{d}_{2}$, where $\mathbf{d}_{1}, \mathbf{d}_{2}\in\mathcal{A}^{M}$ denote two distinct data vectors without rotation. Similar to \cite{FrequencyNonzero_TIT,ref_LTVchannel_book}, the corresponding conditional PEP can be derived as 
\begin{equation}
	P(\mathbf{d}_{1}\rightarrow\mathbf{d}_{2}|\mathbf{h}^{T},\mathbf{d}_{1})=Q\left(\sqrt{\frac{||\mathbf{h}^{T}(\mathbf{X}_{1}-\mathbf{X}_{2})||^{2}}{2N_{0}}}\right),
\end{equation}
where we have $\mathbf{h}=[h_{0},h_{1},\cdots,h_{L-1}]^{T}$ and $\mathbf{X}_{i}=\left[\mathbf{H}_{0}\mathbf{d}_{i},\mathbf{H}_{1}\mathbf{d}_{i},\cdots,\mathbf{H}_{L-1}\mathbf{d}_{i}\right]^{T}$ for $i=1,2$. $Q(x)$ denotes the tail of the standard Gaussian random variable. By adopting the singular value decomposition of $(\mathbf{X}_{1}-\mathbf{X}_{2})(\mathbf{X}_{1}-\mathbf{X}_{2})^{H}$ as $\mathbf{U}\mathbf{\Lambda}\mathbf{U}^{H}$ where $\mathbf{U}$ is a unitary matrix and $\mathbf{\Lambda}=\text{diag}(\lambda_{0}^{2},\lambda_{1}^{2},\cdots,\lambda_{L-1}^{2})$, $||\mathbf{h}^{T}(\mathbf{X}_{1}-\mathbf{X}_{2})||^{2}$ can be simplified as 
\begin{equation}
	||\mathbf{h}^{T}(\mathbf{X}_{1}-\mathbf{X}_{2})||^{2}=\sum_{l=0}^{R-1}\lambda_{l}^{2}|\tilde{h}_{l}|^{2},
\end{equation}
where $R$ denotes the rank of the difference matrix $\mathbf{X}_{1}-\mathbf{X}_{2}$ while $\tilde{h}_{l}\overset{\text{i.i.d.}}{\sim}\mathcal{CN}(0,\frac{1}{P})$ thanks to the unitary property of $\mathbf{U}$. As a result, the average PEP can be derived as
\begin{equation}
	\begin{aligned}
		P(\mathbf{d}_{1}\rightarrow\mathbf{d}_{2})&=\mathbb{E}\left[Q\left(\sqrt{\frac{||\mathbf{h}^{T}(\mathbf{X}_{1}-\mathbf{X}_{2})||^{2}}{2N_{0}}}\right)\right]\\
		&\leq\prod_{l=0}^{R-1}\frac{1}{1+\frac{\lambda_{l}^{2}}{4PN_{0}}}\leq\gamma(\frac{1}{N_{0}})^{-R}.
	\end{aligned}
	\label{PEP_ave}
\end{equation}
From \eqref{PEP_ave}, it can be observed that the exponent of the SNR term is equal to the rank of the difference matrix. In high SNR regions, the PEP with the minimum value dominates the overall error probability. Let $\mathbf{z}=\mathbf{d}_{1}-\mathbf{d}_{2}$ denote the difference vector between two distinct data vectors, the achieved diversity order $G_{d}$ can be derived by
\begin{equation}
	G_{d}=\mathop{\min}_{\mathbf{z}\in\mathcal{B}^{M}\backslash\left\{\mathbf{0}\right\}}\text{rank}\left(\left[\mathbf{H}_{0}\mathbf{z},\mathbf{H}_{1}\mathbf{z},\cdots,\mathbf{H}_{L-1}\mathbf{z}\right]\right),
	\label{rank_begin}
\end{equation} 
where we have\footnote{If the modulation design needs to support $Q$ mapping alphabets as $\mathcal{A}_{1},\mathcal{A}_{1},\cdots\mathcal{A}_{Q}$, we can simply set $\mathcal{B}=\bigcup_{i=1}^{Q}(\mathcal{A}_{i}-\mathcal{A}_{i})=\bigcup_{i=1}^{Q}\{x-y|x, y\in\mathcal{A}_{i}\}$ or even relax as $\mathcal{A}=\bigcup_{i=1}^{Q}\mathcal{A}_{i}$, which will not affect the derivations in this paper since $\mathcal{A}$ still has finite elements.} $\mathcal{B}=\mathcal{A}-\mathcal{A}=\left\{a_{1}-a_{2}|a_{1},a_{2}\in\mathcal{A}\right\}$ to traverse all possible error patterns. On the other hand, since $\mathbf{G}^{-1}$ exists, \eqref{rank_begin} can be transferred as 
\begin{equation}
	G_{d}=\mathop{\min}_{\mathbf{z}\in\mathcal{B}^{M}\backslash\left\{\mathbf{0}\right\}}\text{rank}\left(\left[\tilde{\mathbf{H}}_{0}\mathbf{z},\tilde{\mathbf{H}}_{1}\mathbf{z},\cdots,\tilde{\mathbf{H}}_{L-1}\mathbf{z}\right]\right),
	\label{diversity_rank_moduonly}
\end{equation}
where we have $\tilde{\mathbf{H}}_{l}=\mathbf{A}_{l}\mathbf{\Psi\Phi}$. It indicates that the design of demodulation matrices will not influence the diversity order\footnote{However, the design of demodulation will influence the structure of the equivalent channels. It determines the complexity of fully exploiting the diversity. Therefore, designing appropriate $\mathbf{G}$ is still meaningful.}. Therefore, we aim to find appropriate conditions for modulation schemes to enjoy $\text{rank}\left(\left[\tilde{\mathbf{H}}_{0}\mathbf{z},\tilde{\mathbf{H}}_{1}\mathbf{z},\cdots,\tilde{\mathbf{H}}_{L-1}\mathbf{z}\right]\right)=L$, which means full multipath diversity can be extracted. 
\section{Full Diversity Based on Linearly Precoded CP-OFDM Systems}
\label{sec_spread}
In this section, the linearly precoded CP-OFDM systems are adopted. Let $\mathbf{\Psi}=\mathbf{A}_{\text{CP}}\mathbf{F}_{M}^{H}\tilde{\mathbf{\Psi}}$ denote the equivalent modulation matrix as illustrated in Section \ref{subsec_channel}, where $\mathbf{A}_{\text{CP}}$, $\mathbf{F}_{M}$, and $\tilde{\mathbf{\Psi}}$ represent the CP addition, normalized $M$-point DFT, and frequency-domain linear precoding matrices, respectively. The utilization of CP-OFDM simplifies the analysis significantly, where we present a sufficient condition for $\tilde{\mathbf{\Psi}}$ to achieve full diversity with appropriate constellation rotation in time dispersive channels. Even though the proposed condition is a sufficient one, it is easy to be satisfied in practical generations. The analysis can be extended to doubly dispersive channels by considering the linear spread among multiple transmitted sequences and general physical multipath channels by considering the continuous waveform of CP-OFDM signals.
\subsection{Sufficient Condition for Full Diversity in Time Dispersive Channels}
\label{subsec_spread_delayOnly}
The reason to select linearly precoded CP-OFDM is that the parallel transmission in the frequency domain simplifies the analysis significantly. Let $\mathbf{x}=\tilde{\mathbf{\Psi}}\mathbf{\Phi{d}}$ and $\mathbf{y}=\mathbf{F}_{M}\mathbf{r}$ denote the frequency domain transmitted and received elements, it can be derived as
\begin{equation}
	\mathbf{y}=\mathbf{D}_{H}\mathbf{x}+\mathbf{F}_{M}\mathbf{w},
	\label{IO_OFDM}
\end{equation}
where we have $\mathbf{D}_{H}=\sum_{l=0}^{L-1}h_{l}\mathbf{D}_{l}$ and $\mathbf{D}_{l}=\text{diag}\left(1,e^{-j2\pi\frac{l}{M}},\cdots,e^{-j2\pi\frac{l(M-1)}{M}}\right)$. Based on the characterization in \eqref{IO_OFDM}, the precoding matrix $\tilde{\mathbf{\Psi}}$ can be set to extract full diversity with appropriate phase rotation matrix $\mathbf{\Phi}$. Similar to the analysis in \cite{FrequencyNonzero_first_SPAWC,FrequencyNonzero_TIT,VandermondeLCP_OFDM_delayOnly}, if $\mathbf{x}_{e}=\tilde{\mathbf{\Psi}}\mathbf{\Phi z}$ has no entry equal to zero for $\forall~  \mathbf{z}\in\mathcal{B}^{M}\backslash\left\{\mathbf{0}\right\}$, full diversity order can be achieved. To satisfy this condition, $\tilde{\mathbf{\Psi}}$ should contain no zero entries by considering the scenarios that $\mathbf{z}$ has only one nonzero element. On the other hand, if this condition holds, there are infinite phase rotation choices for $\phi_{0},\phi_{1},\cdots,\phi_{M-1}$ to achieve full diversity, which can be seen clearly in the following theorem.\par 
\begin{theorem}
	\label{th1_spread}
	\rm
	If $\tilde{\mathbf{\Psi}}$ satisfies $|\tilde{\mathbf{\Psi}}(p,q)|>0$ for $\forall p,q$, there are infinite choices for $\phi_{0},\phi_{1},\cdots,\phi_{M-1}$ to achieve full diversity in time dispersive channels. Meanwhile, if $\phi_{0},\phi_{1},\cdots,\phi_{M-1}$ are randomly generated\footnote{Please note that although the angles $\phi_{0},\phi_{1},\cdots,\phi_{M-1}$ for constellation rotation are generated randomly, it is not required to change in each channel/data realization. As a result, it can be easily absorbed into the modulation schemes in practical systems.} from $[0,2\pi)$, the probability that precoding/spread matrix $\mathbf{\Psi\Phi}$ achieves full diversity is 1.
	\begin{IEEEproof}
		The proof is provided in Appendix \ref{th1spread_proof}.
	\end{IEEEproof}
\end{theorem}
Since a random spread matrix $\tilde{\mathbf{\Psi}}$ is almost impossible to have zero entries in practical generations, a phase rotation pattern generated randomly can guarantee the full multipath diversity according to \textbf{Theorem \ref{th1_spread}}. This is different from the prior analysis in \cite{FrequencyNonzero_first_SPAWC,FrequencyNonzero_TIT,phase_rota_DFTsOFDM,VandermondeLCP_OFDM_delayOnly} where very limited spread matrices can be selected like Vandermonde matrices. It can also be extended to the linearly grouping precoded scenarios like \cite{VandermondeLCP_OFDM_delayOnly,group_OFDM}, where the spread matrix for each group satisfies the conditions in \textbf{Theorem \ref{th1_spread}}. By utilizing the grouping, the proposed scheme can support multiuser uplink and downlink transmission with orthogonal multiple access, where users are allocated non-overlapping subcarriers.\par 
\begin{remark}
	\label{remark_DFTs_OFDM}
	\rm
	A typical choice for $\tilde{\mathbf{\Psi}}$ is the $M$-point normalized DFT matrix $\mathbf{F}_{M}$. In this case, we have DFT-s-OFDM systems. However, the diversity order of DFT-s-OFDM is only $1$, which can be verified simply by utilizing $\mathbf{d}_{1}(q)=1$ and $\mathbf{d}_{2}(q)=-1$ for $\forall q$ in Section \ref{subsec_diversity}. This conclusion has also been provided in \cite{CP_SC_diversity1}. On the other hand, DFT-s-OFDM with a random phase rotation pattern can achieve full multipath diversity over time dispersive channels according to \textbf{Theorem \ref{th1_spread}}, which can also be demonstrated by numerical results in Section \ref{sec_simu}. Please note that it is different from \cite{phase_rota_DFTsOFDM} because the phase rotation needs to be acquired via the computer search in \cite{phase_rota_DFTsOFDM}. However, in this paper, the rotation angles can be randomly generated over an arbitrary continuous distribution over $[0,2\pi)$ according to the analysis in \textbf{Theorem \ref{th1_spread}}.
\end{remark}
\subsection{Extension and Discussions}
\label{subsec_spread_discussion}
In this subsection, we extend the discussions for linearly precoded/spread CP-OFDM systems to more general scenarios. At first, the following corollary can be derived by considering general physical channels in \eqref{LTI_continuous} and the continuous waveform of CP-OFDM systems.\par 
\begin{corollary}
	\rm
	Consider the linearly precoded continuous CP-OFDM systems and the general time dispersive channels in \eqref{LTI_continuous}, where the precoding/spread matrix $\tilde{\mathbf{\Psi}}$ satisfies $\tilde{\mathbf{\Psi}}(p,q)\ne0$ for $\forall p,q$ and the rotation angles $\phi_{0},\phi_{1},\cdots,\phi_{M-1}$ are randomly generated from $[0,2\pi)$. The spread by $\tilde{\mathbf{\Psi}}\mathbf{\Phi}$ will achieve the full multipath diversity order of $I$ in time dispersive channels in \eqref{LTI_continuous} if $\tau_{i}\ne\tau_{i^{\prime}}$ for $\forall i\ne i^{\prime}$ and $\tau_{i}\Delta{f}<1$, where $\Delta{f}$ denotes the subcarrier spacing.\par 
	\begin{IEEEproof}
		The continuous waveform of CP-OFDM can be derived as $s(t)=\sum_{m=0}^{M-1}\mathbf{x}[m]e^{j2\pi m\Delta{f}t}$ for $-T_{\text{CP}}<t<T$, where $T$, $T_{\text{CP}}$ and $\Delta{f}=\frac{1}{T}$ denote the OFDM symbol duration, CP duration and subcarrier spacing, respectively. By repeating the analysis in Section \ref{subsec_diversity}, we can then deduce that if $\left[\mathbf{D}_{1}\mathbf{x}_{e},\mathbf{D}_{2}\mathbf{x}_{e},\cdots,\mathbf{D}_{I}\mathbf{x}_{e}\right]$ has full rank, the full diversity order of $I$ can be achieved, where we have $\mathbf{D}_{i}=\text{diag}\left(1,e^{-j2\pi \Delta{f}\tau_{i}},\cdots,e^{-j2\pi(M-1)\Delta{f}\tau_{i}}\right)$. Considering the full rank of Vandermonde matrices, if $\mathbf{x}_{e}=\tilde{\mathbf{\Psi}}\mathbf{\Phi z}$ has no entry equal to zero for $\forall  \mathbf{z}\in\mathcal{B}^{M}\backslash\left\{\mathbf{0}\right\}$, full diversity order can be achieved, which completes the proof by employing \textbf{Theorem \ref{th1_spread}}.
	\end{IEEEproof}
\end{corollary}
This corollary also indicates the robustness of DFT-s-OFDM systems with random phase rotation when it comes to the diversity order over general physical channels. For doubly selective fading scenarios, similar derivations can be attained by considering the joint precoder among $N$ OFDM symbols. Let $Y(n,m)$ and $X(n,m)$ represent the component at the $n$-th OFDM symbol and $m$-th subcarrier, the input-output relation for CP-OFDM systems over doubly selective channels can be approximately derived as \cite{noICI_TVOFDM,noISI_TIT}
\begin{equation}
	Y(n,m)\approx\sum_{i=1}^{I}\tilde{h}_{i}e^{j2\pi\left(\nu_{i}n(T+T_{\text{CP}})-\tau_{i}m\Delta{f}\right)}X(n,m)+W(n,m),
	\label{IO_double_OFDM_noICI}
\end{equation}
where $\nu_{i}$ is the Doppler shift associated with the $i$-th channel path while $W(n,m)$ represents the AWGN samples. Full diversity can still be achieved if the precoded results for each pair of distinct data vectors spread all time-frequency grids. As a result, similar precoder configurations can be selected as in \textbf{Theorem \ref{th1_spread}}. However, this approximation in \eqref{IO_double_OFDM_noICI} ignores the inter-carrier interference (ICI) in OFDM systems brought by the time-variant channels, which indicates the spread might be not appropriate for high Doppler spread. To address this issue, a more comprehensive analysis will be depicted in Section \ref{subsec_modulation_discussion} by focusing on the direct modulation design, where we do not assume ICI-free characterization in the frequency domain.\par  
\section{Full Diversity Based on General Linear Modulation Systems}
\label{sec_modulation}
In this section, we focus on the design criteria for general linear modulation denoted by $\mathbf{\Psi}$ to achieve full multipath diversity, where $\mathbf{\Psi}$ can be selected arbitrarily from $\mathbb{C}^{(M+M_{p})\times{M}}$ rather than derived by CP-OFDM modulation with a linear precoder. We first propose the sufficient and necessary condition for $\mathbf{\Psi}$ to enjoy full multipath diversity in time dispersive channels by enabling appropriate constellation rotation, where we also prove that a random phase rotation can reach full diversity with the probability of $1$. The extension to doubly dispersive channels is then developed via similar derivations, where the ICI-free assumption in the frequency domain is not required to better match the realistic doubly dispersive channels.\par 
\subsection{Sufficient and Necessary Condition for Full Diversity in Time Dispersive Channels}
\label{subsec_modulation_delayOnly}
In this subsection, the sufficient and necessary condition for $\mathbf{\Psi}$ to reach maximum diversity, i.e., full rank in \eqref{diversity_rank_moduonly}, is presented. Before the derivations, a lemma in linear algebra is established to serve as the basis for the following illustrations.
\begin{lemma}
	\label{lemma1_lowrank_finite}
	\rm
	Let $\mathbf{A},\mathbf{B}\in\mathbb{C}^{m\times{n}}$ denote 2 matrices with full rank, i.e., $\text{rank}(\mathbf{A})=\text{rank}(\mathbf{B})=\min(m,n)$. There are at most $\max(m,n)$ distinct values of complex number $c$ to satisfy $\text{rank}(\mathbf{A}+c\mathbf{B})<\min(m,n)$.
	\begin{IEEEproof}
		The proof is provided in Appendix \ref{lemma1_finite_lowrank_proof}.
	\end{IEEEproof}
\end{lemma}
In the meantime, the full rank should first be guaranteed when $\mathbf{z}$ reveals the one-hot property similar to the illustrations in Section \ref{subsec_spread_delayOnly}, i.e., the number of nonzero elements in $\mathbf{z}$ is only one. By substituting this into \eqref{diversity_rank_moduonly}, it can be deduced that the following equation should be satisfied for $\forall~q=0,1,\cdots,M-1$ as
\begin{equation}
	\text{rank}\left(\left[\mathbf{A}_{0}\mathbf{\Psi}_{q},\mathbf{A}_{1}\mathbf{\Psi}_{q},\cdots,\mathbf{A}_{L-1}\mathbf{\Psi}_{q}\right]\right)=L,
	\label{fullrank_modu_judge1}
\end{equation} 
where $\mathbf{\Psi}_{q}$ represents the $q$-th column vector of the modulation matrix $\mathbf{\Psi}$. Let $\mathbf{J}_{q}=\left[\mathbf{A}_{0}\mathbf{\Psi}_{q},\mathbf{A}_{1}\mathbf{\Psi}_{q},\cdots,\mathbf{A}_{L-1}\mathbf{\Psi}_{q}\right]$ denote the judgment matrix in \eqref{fullrank_modu_judge1} to determine the PEP when the detection error only occurs at the $q$-th data symbol. $\mathbf{J}_{q}\in\mathbb{C}^{M\times{L}}$ can be equivalently derived as 
\begin{equation}
	\mathbf{J}_{q}(p,l)=\mathbf{\Psi}(p-l,q).
	\label{judgematrix_modu}
\end{equation}
It is apparent that a necessary condition for full diversity is that $\text{rank}(\mathbf{J}_{q})=L$ for $\forall q=0,1,\cdots,M-1$. However, the following theorem demonstrates that it is also a sufficient condition for achieving maximum multipath diversity.
\begin{theorem}
	\label{th2_modu}
	\rm 
	Phase rotation patterns to achieve full diversity in time dispersive channels exist if and only if $\text{rank}(\mathbf{J}_{q})=L$ for $\forall q=0,1,\cdots,M-1$. Besides, there are infinite phase choices and if $\phi_{0},\phi_{1},\cdots,\phi_{M-1}$ are selected randomly from $[0,2\pi)$, maximum diversity can be attained with the probability of $1$.
	\begin{IEEEproof}
		The proof is provided in Appendix \ref{th2_modu_proof}.
	\end{IEEEproof}
\end{theorem}
\textbf{Theorem \ref{th2_modu}} indicates that to achieve full diversity, the initial problem in \eqref{diversity_rank_moduonly} can be divided into 2 sub-problems. The designing criteria for $\mathbf{\Psi}$ can then be concentrated for the PEP analysis when there is only one nonzero element in the error vector $\mathbf{z}$, which is a simpler task since it is only determined by the property of each column of $\mathbf{\Psi}$. The other part is to obtain a constellation rotation pattern randomly, which also involves slight complexity in practical generations. Meanwhile, for CP-enabled single carrier systems, full diversity can also be obtained by random phase rotation patterns by checking $\mathbf{J}_{q}$, which corresponds to the conclusions in \textbf{Remark \ref{remark_DFTs_OFDM}}. Furthermore, it leads to the following corollary for a more general diversity analysis. 
\begin{corollary}
	\label{corollary_diversity_general}
	\rm
	If the first $l$ columns of $\mathbf{J}_{q}$ are linearly independent for $\forall q=0,\cdots,M-1$, the diversity order of $l$ can be obtained via a phase rotation pattern generated randomly.
	\begin{IEEEproof}
		The proof can be achieved by a similar proof in Appendix \ref{th2_modu_proof} for the first $l$ columns. 
	\end{IEEEproof}
\end{corollary}\par 
Based on the aforementioned analysis, a direct corollary can also be deduced to indicate the lower bound of the diversity order for arbitrary modulation designs. 
\begin{corollary}
	\rm
	If the last prefix element, i.e., the $M_{p}$-th row of $\mathbf{\Psi}$ can be adjusted, arbitrary modulation design can achieve the diversity order of at least $2$.
	\begin{IEEEproof}
		Consider the first 2 columns of $\mathbf{J}_{q}$ as
		\begin{equation}
			\mathbf{J}_{q}^{2}=
			\begin{pmatrix}
				\mathbf{\Psi}_{q}(0) &\mathbf{\Psi}_{q}(-1)\\
				(\tilde{\mathbf{\Psi}}_{q})_{-0}
				&(\tilde{\mathbf{\Psi}}_{q})_{-(M-1)}
			\end{pmatrix},
		\end{equation}
		where $\tilde{\mathbf{\Psi}}_{q}$ denotes the last $M$ rows of $\mathbf{\Psi}_{q}$ while $\mathbf{u}_{-n}$ is $\mathbf{u}$ without the $n$-th entry for a column vector $\mathbf{u}$. If $\mathbf{\Psi}_{q}(0)=0$, any nonzero value of $\mathbf{\Psi}_{q}(-1)$ will make $\text{rank}(\mathbf{J}_{q}^{2})=2$. If $\mathbf{\Psi}_{q}(0)\ne0$, a natural choice is to set
		\begin{equation}
			\bigg(\mathbf{\Psi}_{q}(0)\bigg)^{*}\mathbf{\Psi}_{q}(-1)+\bigg((\tilde{\mathbf{\Psi}}_{q})_{-0}\bigg)^{H}(\tilde{\mathbf{\Psi}}_{q})_{-(M-1)}=0,
		\end{equation}
		which assures the full rank of $\mathbf{J}_{q}^{2}$ thanks to the property of orthogonality. As a result, $\mathbf{J}_{q}^{2}$ has the rank of $2$ by selecting appropriate value for $\mathbf{\Psi}_{q}(-1)$. As a result, arbitrary modulation design can achieve the diversity order of at least $2$ if the last prefix element can be adjusted, which completes the proof by utilizing \textbf{Corollary \ref{corollary_diversity_general}}.
	\end{IEEEproof}
\end{corollary}
\subsection{Extension and Discussions}
\label{subsec_modulation_discussion}
Similar to the generalization in Section \ref{subsec_spread_discussion}, \textbf{Theorem \ref{th2_modu}} can also be extended to doubly-dispersive scenarios in \eqref{IO_double}. Let $\mathbf{V}_{q}\in\mathbb{C}^{M\times L(2K+1)}$ represent the judgment matrix for each column of the modulation matrix, which can be derived as
\begin{equation}
	\mathbf{V}_{q}(p,kL+l)=e^{j2\pi\frac{kp}{M}}\mathbf{\Psi}(p-l,q).
\end{equation}
The following corollary can then be established by a similar proof in Appendix \ref{th2_modu_proof}.
\begin{corollary}
	\rm 
	\label{corollary_doubly_general}
	Phase rotation patterns to achieve full diversity exist if and only if $\text{rank}(\mathbf{V}_{q})=L(2K+1)$ for $\forall q=0,1,\cdots,M-1$. Besides, there are infinite phase choices and if $\phi_{0},\phi_{1},\cdots,\phi_{M-1}$ are selected randomly from $[0,2\pi)$, maximum diversity can be attained with the probability of $1$.
\end{corollary}
This corollary also explains why the emerging modulation techniques, such as OTFS \cite{OTFS_diversity_TWC,OTFS_diversity_WCL,OTFS_diversity_TVT,shuangyang_OTFS}, ODDM \cite{ref_ODDM}, and AFDM \cite{ref_AFDM} modulation can be treated as full-diversity schemes. The connection lies in that if the impact of nonsingular demodulation is taken into consideration, there is only one nonzero element in each column of the judgment matrix for PEP analysis when the number of error symbols is one, and their locations among columns do not overlap. As a result, $\text{rank}(\mathbf{V}_{q})=L(2K+1)$ can be guaranteed and all of these schemes can achieve the full diversity with random phase rotation patterns. Actually, traditional OTFS and ODDM systems do not enjoy full diversity in all scenarios. Only when the phase rotation is appropriately considered, these schemes can obtain maximum multipath diversity, i.e., separability leads to full diversity order, which has also been embodied in the proof for AFDM schemes in \cite{ref_AFDM}. \par 
\begin{figure}
	\centering{\includegraphics[width=0.9\linewidth]{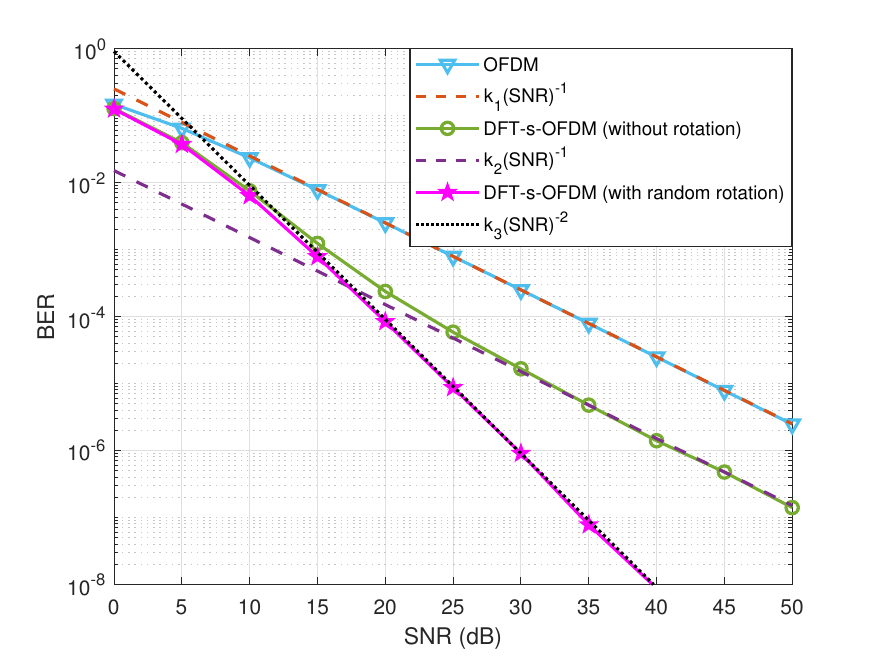}}
	\caption{BER against SNR under $M=4$ and time dispersive channels in \eqref{IO_delayOnly} with $L=2$.}
	\label{SimuFig_L2_M4}	
\end{figure}
In fact, the most meaningful remark from this section is that by involving the random phase rotation, the design criteria for modulation schemes can be restricted on the one-hot scenarios even though full diversity is desired. It also reduces the complexity of verifying the full diversity with a pre-defined modulation strategy significantly since the rank of only $M$ matrices is required to check rather than $|\mathcal{B}|^{M}-1$. \par 
It is also worth pointing out that the realistic performance also depends on the receiver scheme since ML involves prohibitive computational complexity, i.e., even though two modulation schemes have different diversity orders, their reliability might be similar if other detectors are deployed rather than the ML-based ones. For example, let us consider the linear zero-forcing (LZF) equalization followed by the hard decision function as the receiver, where two modulation strategies only differ in the phase rotation pattern. The transmission model can then be derived as $\mathbf{y}_{1}=\mathbf{H\mathbf{\Phi}_{1}}\mathbf{d}+\mathbf{w}$ and $\mathbf{y}_{2}=\mathbf{H\mathbf{\Phi}_{2}}\mathbf{d}+\mathbf{w}$, respectively, where $\mathbf{H}$ denotes the equivalent channel matrix considering the modulation matrix. Let $\hat{\mathbf{d}}_{1}$ and $\hat{\mathbf{d}}_{2}$ denote the equalized results as
\begin{equation}
	\begin{cases}
		\hat{\mathbf{d}}_{1}=(\mathbf{H\Phi}_{1})^{-1}\mathbf{y}_{1}=\mathbf{d}+\mathbf{\Phi}_{1}^{H}\tilde{\mathbf{w}}\\
		\hat{\mathbf{d}}_{2}=(\mathbf{H\Phi}_{2})^{-1}\mathbf{y}_{2}=\mathbf{d}+\mathbf{\Phi}_{2}^{H}\tilde{\mathbf{w}}\\
	\end{cases},
\end{equation}
where we have $\tilde{\mathbf{w}}=\mathbf{H}^{-1}\mathbf{w}$. It can then be derived that $\hat{\mathbf{d}}_{i}(q)=\mathbf{d}(q)+\mathbf{w}_{i}(q)$ for $i=1,2$, where we have $\mathbb{E}(\mathbf{w}_{i}(q))=0$ and $\mathbb{E}(|\mathbf{w}_{i}(q)|^{2})=\mathbb{E}(|\tilde{\mathbf{w}}(q)|^{2})$. It indicates that $\mathbf{w}_{1}(q)$ and $\mathbf{w}_{2}(q)$ follow the same complex Gaussian distribution. Therefore, these two schemes lead to the same error performance if the LZF detector is utilized. However, as indicated before, they may have different diversity orders if the ML detector is employed, which shows that optimizing the diversity order under ML assumptions might be not sufficient for performance promotion under realistic receiver designs.\par 

\section{Numerical Results}
\label{sec_simu}
In this section, numerical results are presented to illustrate the performance of various modulation schemes and compare with the analysis in this paper. To reduce the complexity overload and verify the analysis in practical generations, the rotation angles $\phi_{0},\phi_{1},\cdots,\phi_{M-1}$ are randomly and independently generated over $[0,2\pi)$. We first present the results under time dispersive channels and then depict the extensions to doubly dispersive channels. According to \textbf{Remark \ref{remark_DFTs_OFDM}} and the corresponding illustrations in Section \ref{sec_modulation}, we select the conventional DFT-s-OFDM and OFDM as the comparison with the diversity order of $1$. Finally, binary phase shift keying (BPSK) is selected as the mapping alphabet and the ML detector is utilized to exploit the diversity.\par 
Fig. \ref{SimuFig_L2_M4} presents BER against SNR under $M=4$ and time dispersive channels in \eqref{IO_delayOnly} with $L=2$. It is obvious that the diversity order of both OFDM and DFT-s-OFDM systems is $1$, which is embodied as the slope at the high SNR regime. By appending the phase rotation generated randomly at the beginning of numerical experiments, a full diversity of $2$ is achieved. Therefore, when SNR$>$15 dB, the BER of DFT-s-OFDM systems with phase rotation is much lower than that without rotation. When BER is about $10^{-7}$, the SNR precedence of more than $15$ dB can be attained by the constellation rotation. The results also indicate that separability does not provide full diversity even though it brings significant convenience by enabling a phase rotation pattern, which corresponds to the analysis in \textbf{Theorems \ref{th1_spread} and \ref{th2_modu}} in this paper.\par  
\begin{figure}
	\centering{\includegraphics[width=0.9\linewidth]{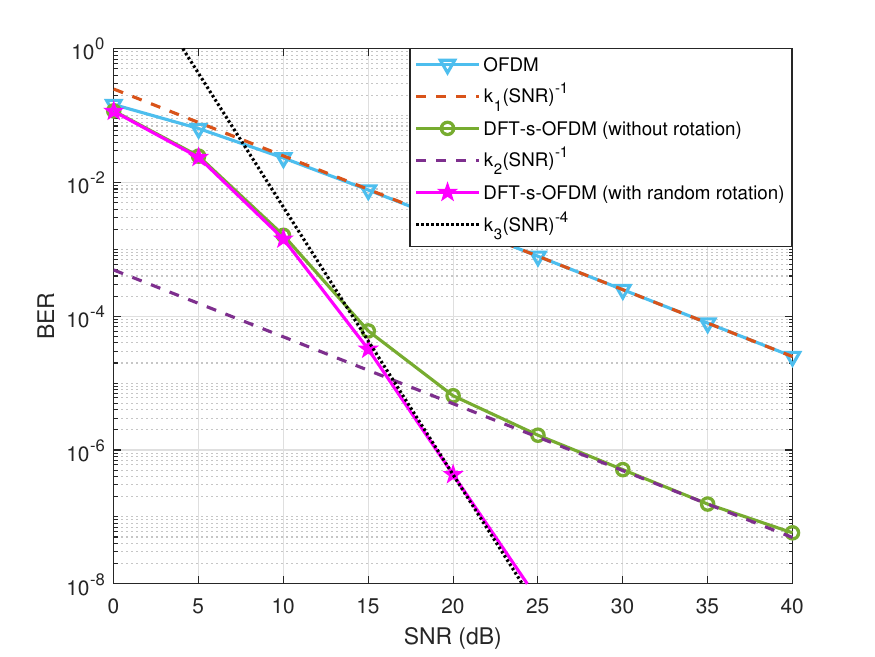}}
	\caption{BER against SNR under $M=8$ and time dispersive channels in \eqref{IO_delayOnly} with $L=4$.}
	\label{SimuFig_L4_M8}	
\end{figure} 
Fig. \ref{SimuFig_L4_M8} depicts BER against SNR under $M=8$ and time dispersive channels in \eqref{IO_delayOnly} with $L=4$. It is obvious that DFT-s-OFDM systems with random phase rotation still reveal a full diversity order of $4$. Similar to the results in Fig. \ref{SimuFig_L2_M4}, the performance superiority brought by the constellation rotation begins at about SNR=15 dB. However, compared with Fig. \ref{SimuFig_L2_M4}, the diversity one regime starts at a much lower BER value of about $10^{-6}$, which corresponds to the simulations in \cite{OTFS_diversity_TWC}. Meanwhile, though conventional DFT-s-OFDM and OFDM systems share the same diversity order of $1$, the performance gap is larger than $25$ dB. It indicates that the coding gain of DFT-s-OFDM systems outperforms OFDM schemes significantly. Also, since each data symbol can fully utilize the frequency domain gains to enhance the reliability, DFT-s-OFDM outperforms OFDM even though in the low-SNR regions. At last, in the finite SNR regime, the diversity order can be larger than 1 for DFT-s-OFDM systems without phase rotation before the diversity one regime takes over. It is noteworthy that the BER curve of DFT-s-OFDM displays a tri-stage characteristic with a turning point at SNR=20 dB. This behavior is attributed to the fact that the maximum diversity in PEPs of DFT-s-OFDM is 4, which can be achieved by the PEPs associated with single-symbol error events. Across all PEPs, those with larger diversity gains generally have smaller coding gains, whereas cases with smaller diversity gains tend to possess larger coding gains. When SNR is low, the high-diversity components prevail, resulting in a higher BER slope. As the SNR increases beyond a critical level, the low-diversity components become predominant, leading to a noticeable reduction in slope.\par 
\begin{figure}
	\centering{\includegraphics[width=0.9\linewidth]{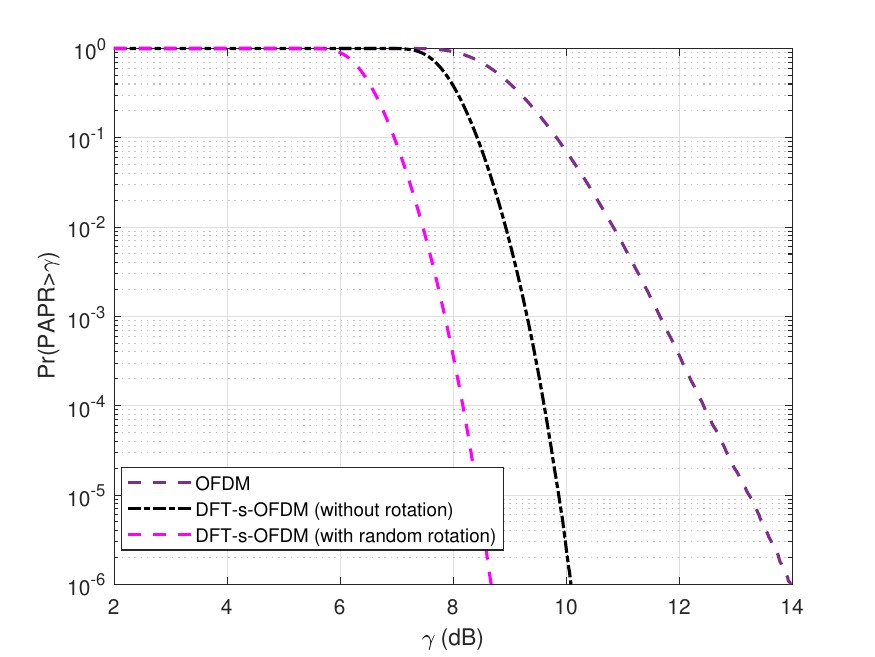}}
	\caption{CCDF evaluation with the phase rotation. }
	\label{SimuFig_PAPR}	
\end{figure}
To provide a comprehensive evaluation, Fig. \ref{SimuFig_PAPR} presents the peak-to-average power ratio (PAPR) by plotting the complementary cumulative distribution function (CCDF) under $M=1024$. The oversampling factor is set as $8$ to better approximate the realistic PAPR values \cite{ref_PAPR_oversampling}. It is obvious that the DFT-s-OFDM systems with random constellation rotation enjoy the lowest PAPR. In fact, the rotation has been treated as a technique to reduce PAPR in DFT-s-OFDM systems \cite{PAPR_rota_ref}. By utilizing the random rotation, more than 1 dB reduction of PAPR can be achieved compared with conventional DFT-s-OFDM systems with the CCDF threshold of $10^{-4}$.\par 
\begin{figure}
	\centering{\includegraphics[width=0.9\linewidth]{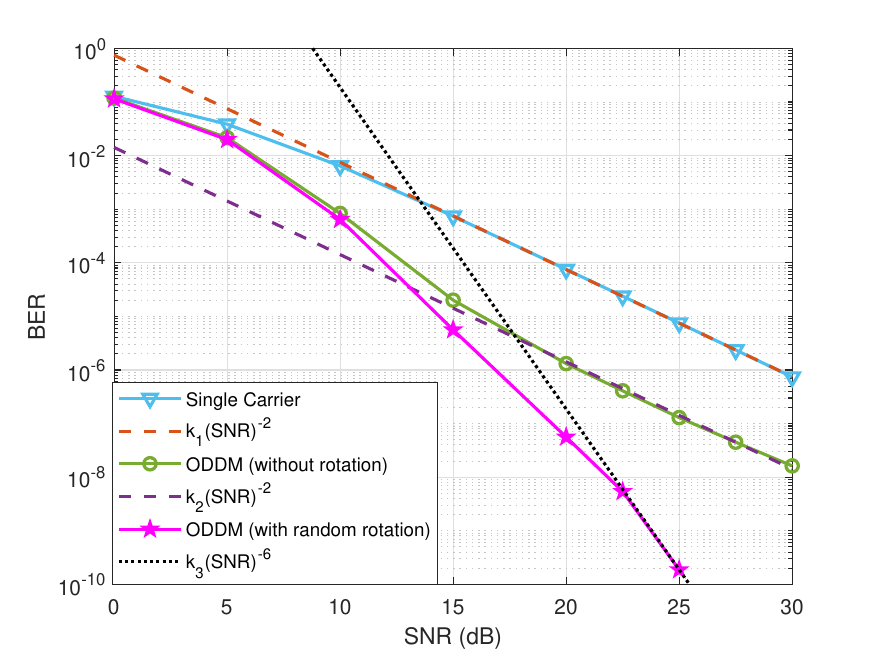}}
	\caption{BER against SNR under $M=8$ and doubly dispersive channels in \eqref{IO_double} with $K=1$ and $L=2$.}
	\label{SimuFig_L6_double}	
\end{figure} 
To demonstrate the analysis under doubly dispersive channels in Section \ref{subsec_modulation_discussion}, we carry out the BER test under $M=8$ and ML detection in doubly dispersive channels in Fig. \ref{SimuFig_L6_double}. For the channel realizations, we set $K=1$ and $L=2$ as the maximum Doppler and delay tap as in \cite{OTFS_diversity_TWC,OTFS_diversity_WCL}. The number of multicarrier symbols and subcarriers in ODDM systems is set as $2$ and $4$ \cite{OTFS_diversity_TWC,OTFS_diversity_WCL}, respectively. Due to the separability based on the input-output analysis in \cite{OTFS_diversity_WCL}, the condition in \textbf{Corollary \ref{corollary_doubly_general}} is satisfied. Therefore, the ODDM modulation with random constellation rotation achieves the maximum diversity order of $L(2K+1)=6$. In contrast, the diversity of conventional ODDM modulation is only $2$ even though ML detection is deployed, which corresponds to the analysis in \cite{OTFS_diversity_TWC,OTFS_diversity_WCL,OTFS_diversity_TVT}. The BER curve of ODDM modulation also exhibits a three-stage property with the turning point at SNR=20 dB, which is similar to DFT-s-OFDM systems in time dispersive channels in Fig. \ref{SimuFig_L4_M8}.\par 
\begin{figure}
	\centering{\includegraphics[width=0.9\linewidth]{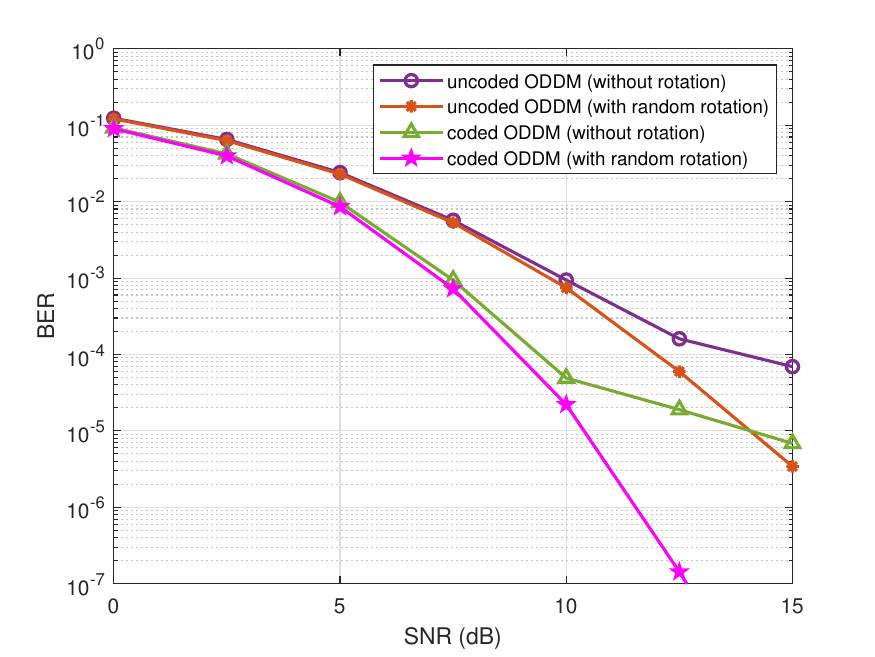}}
	\caption{BER against SNR under practical detection and channels.}
	\label{SimuFig_practical}	
\end{figure} 
Finally, the merit of the proposed analysis is confirmed by plotting BER against SNR for both uncoded and coded ODDM systems under practical assumptions in Fig. \ref{SimuFig_practical}. The number of multicarrier symbols and subcarriers in ODDM systems is set as $512$ and $16$, respectively. The carrier frequency is $5$ GHz while the delay resolution is $0.13~\mu{s}=\frac{1}{512\times15~\text{kHz}}$. The 5G New Radio (NR) low-density parity-check (LDPC) code is adopted with the coding rate of $\frac{3}{4}$ \cite{ref_LDPC}. For channel realizations, we adopt the TDLC300 delay-power profile \cite{ref_TDL}, where each path has a randomly generated Doppler shift according to Jakes' formula and a maximum velocity of 360 km/h. The approximate message passing (AMP) detection \cite{ref_AMP} is utilized. It is obvious that thanks to the random constellation rotation, lower BER can be observed for both coded and uncoded scenarios. When SNR=15 dB, uncoded BER with random rotation is less than $4\times10^{-6}$ while that without rotation is about $7\times10^{-5}$. Similar phenomena can also be observed for coded transmission. When SNR is 12.5 dB, the coded BER with random constellation approaches $10^{-7}$, whereas the scheme without phase rotation remains above $10^{-5}$. Besides, ODDM systems with the random constellation rotation exhibit a significantly steeper decay slope for both coded and uncoded scenarios in high-SNR regions. This corresponds to the diversity analysis in this paper. In conclusion, even though the optimal diversity order cannot be achieved considering practical detectors, performance precedence can still be obtained by adopting the random constellation, which further demonstrates the value of this work.\par 
\section{Conclusions}
\label{sec_conclusion}
In this paper, we developed the random constellation rotation to ease the modulation design for full multipath diversity in dispersive channels. We first investigated the linearly precoded CP-OFDM systems and proposed a sufficient but easily satisfied condition for the precoding matrix to extract full diversity order. We then derived the sufficient and necessary condition for general modulation schemes to achieve the same target. We demonstrated that both schemes satisfying the proposed conditions enjoy the probability of $1$ to have full diversity when the rotation angles are generated randomly and the discussions can be extended to doubly dispersive channels for both systems. We also proved that separability itself does not mean full diversity if the phase rotation is absent, which better explains the superiority of the emerging modulation schemes such as OTFS, ODDM, and AFDM. Numerical results were provided at last to verify the derivations in this paper as well. The analysis in this paper could help the research community gain a better understanding of the relationship between multipath diversity and modulation designs from a theoretical perspective. It is also beneficial for inspiring novel modulation designs for emerging applications in next-generation wireless communication systems where full multipath diversity is expected and the corresponding performance analysis can be focused on the PEP investigation with one error data symbol.   
\appendices
\section{Proof of Theorem \ref{th1_spread}}
\label{th1spread_proof}
To prove this theorem, we first provide a construction algorithm for $\phi_{0},\phi_{1},\cdots,\phi_{M-1}$ and then illustrate why the random generation is almost always optimal. Please notice that the objective has been transferred into that $\mathbf{x}_{e}=\tilde{\mathbf{\Psi}}\mathbf{\Phi z}$ has no entry equal to zero for $\forall  \mathbf{z}\in\mathcal{B}^{M}\backslash\left\{\mathbf{0}\right\}$.\par
To realize full diversity based on a pre-defined $\tilde{\mathbf{\Psi}}$, we propose to set the rotation angles from $\phi_{0}$ to $\phi_{M-1}$. At first, $\phi_{0}$ can be randomly selected. Assuming that $\phi_{0},\phi_{1},\cdots\phi_{q-1}$ have been decided, the selection of $\phi_{q}$ should guarantee that $\mathbf{x}_{e}=\tilde{\mathbf{\Psi}}\mathbf{\Phi z}$ has no zero entries for $\forall\mathbf{z}\in\mathcal{B}^{M}\backslash\left\{\mathbf{0}\right\}$ with $\mathbf{z}(q^{\prime})=0$ for $q^{\prime}>q$. To achieve this, the following conditions should be satisfied as
\begin{equation}
	\tilde{\mathbf{\Psi}}(m,q)e^{j\phi_{q}}\mathbf{z}(q)+\sum_{p=0}^{q-1}\tilde{\mathbf{\Psi}}(m,p)e^{j\phi_{p}}\mathbf{z}(p)\ne0,
\end{equation}
for $m=0,1,\cdots,M-1$ and $\mathbf{z}(q)\ne0$. Since $\tilde{\mathbf{\Psi}}(m,q)\ne0$, it can then be transferred as
\begin{equation}
	e^{j\phi_{q}}\ne-\frac{\sum_{p=0}^{q-1}\tilde{\mathbf{\Psi}}(m,p)e^{j\phi_{p}}\mathbf{z}(p)}{\tilde{\mathbf{\Psi}}(m,q)\mathbf{z}(q)}.
	\label{nonzero_phase_neq}
\end{equation}
To select $\phi_{q}$, there are $M|\mathcal{B}|^{q}(|\mathcal{B}|-1)$ inequalities as shown in \eqref{nonzero_phase_neq} to be satisfied. Meanwhile, each inequality in \eqref{nonzero_phase_neq} will exclude at most a possible selection for $\phi_{q}$ in $[0,2\pi)$. Therefore, $\phi_{q}$ can be set as $\phi_{q}\in[0,2\pi]\backslash\mathcal{C}_{q}$, where $\mathcal{C}_{q}$ contains at most $M(|\mathcal{B}|^{q}-1)(|\mathcal{B}|-1)$ elements determined by changing $\mathbf{z}(p)$ for $p=0,1,\cdots,q$ and the index $m$. After $\phi_{M-1}$ is decided, we have
\begin{equation}
	\sum_{q=0}^{M-1}\tilde{\mathbf{\Psi}}(m,q)e^{j\phi_{q}}\mathbf{z}(q)\ne0,
\end{equation}
for $\forall\mathbf{z}\in\mathcal{B}^{M}\backslash\left\{\mathbf{0}\right\}$, which indicates that full multipath diversity can be achieved.\par 
\begin{figure*}
	\begin{equation}
		\label{fullrank_proof_step1_divide}
		\begin{aligned}
			&\left[\tilde{\mathbf{H}}_{0}\mathbf{z},\tilde{\mathbf{H}}_{1}\mathbf{z},\cdots,\tilde{\mathbf{H}}_{L-1}\mathbf{z}\right]\\
			&=\left[\mathbf{A}_{0}\left(\sum_{i=0}^{q}\mathbf{z}(i)e^{j\phi_{i}}\mathbf{\Psi}_{i}\right),\mathbf{A}_{1}\left(\sum_{i=0}^{q}\mathbf{z}(i)e^{j\phi_{i}}\mathbf{\Psi}_{i}\right),\cdots,\mathbf{A}_{L-1}\left(\sum_{i=0}^{q}\mathbf{z}(i)e^{j\phi_{i}}\mathbf{\Psi}_{i}\right)\right]\\
			&=\underbrace{\left[\mathbf{A}_{0}\left(\sum_{i=0}^{q-1}\mathbf{z}(i)e^{j\phi_{i}}\mathbf{\Psi}_{i}\right),\mathbf{A}_{1}\left(\sum_{i=0}^{q-1}\mathbf{z}(i)e^{j\phi_{i}}\mathbf{\Psi}_{i}\right),\cdots,\mathbf{A}_{L-1}\left(\sum_{i=0}^{q-1}\mathbf{z}(i)e^{j\phi_{i}}\mathbf{\Psi}_{i}\right)\right]}_{\mathbf{B}_{\mathbf{z},q-1}}\\
			&+e^{j\phi_{q}}\underbrace{\left[\mathbf{z}(q)\mathbf{A}_{0}\mathbf{\Psi}_{q},\mathbf{z}(q)\mathbf{A}_{1}\mathbf{\Psi}_{q},\cdots,\mathbf{z}(q)\mathbf{A}_{L-1}\mathbf{\Psi}_{q}\right]}_{\mathbf{C}_{\mathbf{z},q}}\\
			&=\mathbf{B}_{\mathbf{z},q-1}+e^{j\phi_{q}}\mathbf{C}_{\mathbf{z},q}
		\end{aligned}
	\end{equation}
	\hrulefill
\end{figure*}
\begin{algorithm}
	\renewcommand{\algorithmicrequire}{\textbf{Input:}}
	\renewcommand{\algorithmicensure}{\textbf{Output:}}
	\caption{Selection of rotation pattern with strictly full multipath diversity}
	\label{alg_rotation}
	\begin{algorithmic}[1]
		\REQUIRE
		$\mathcal{B}$, $\tilde{\mathbf{\Psi}}$, $M$ 
		\ENSURE
		The rotation pattern $\phi_{0},\phi_{1},\cdots,\phi_{M-1}$
		\STATE
		Randomly generate $\phi_{0}$
		\FOR{$q=1:M-1$}
		\STATE
		$\mathcal{C}_{q}=\emptyset$
		\STATE
		Calculate all inadmissible values of $\phi_{q}$ according to \eqref{nonzero_phase_neq} and place them into $\mathcal{C}_{q}$
		\STATE
		Randomly generate $\phi_{q}\in[0,2\pi)\backslash\mathcal{C}_{q}$
		\ENDFOR
	\end{algorithmic}		
\end{algorithm}
Therefore, the construction algorithm for $\phi_{0},\phi_{1},\cdots,\phi_{M-1}$ can be summarized as \textbf{Algorithm \ref{alg_rotation}}, where strictly full diversity order can be achieved. However, to determine $\phi_{q}$, there are $M|\mathcal{B}|^{q}(|\mathcal{B}|-1)$ inequalities to be computed as shown in \eqref{nonzero_phase_neq}. The number of total inequalities can then be derived as
\begin{equation}
	\begin{aligned}
		\sum_{q=1}^{M-1}M|\mathcal{B}|^{q}(|\mathcal{B}|-1)=M(|\mathcal{B}|^{M}-|\mathcal{B}|),
	\end{aligned}
\end{equation}  
which leads to unbearable computational load in practical generations even for small values of $M$ and $|\mathcal{A}|$. For example, more than $5.2\times10^{4}$ inequalities should be calculated for $M=8$ and BPSK alphabets with $|\mathcal{B}|=3$. Fortunately, since $\mathcal{C}_{q}$ contains at most $M|\mathcal{B}|^{q}(|\mathcal{B}|-1)$ elements, there are infinite choices for deciding each rotation angle during the construction, which leads to infinite choices for $\phi_{0},\phi_{1},\cdots,\phi_{M-1}$. The finite nature of $\mathcal{C}_{q}$ also indicates that if $\phi_{q}$ is randomly drawn from a continuous distribution over $[0,2\pi)$, the probability that all inequalities derived from \eqref{nonzero_phase_neq} hold is $1$. Then the proof of \textbf{Theorem \ref{th1_spread}} is completed.\par 
%
\section{Proof of Lemma \ref{lemma1_lowrank_finite}}
\label{lemma1_finite_lowrank_proof}
Because it is equivalent to derive this lemma for $\mathbf{A}^{T}$ and $\mathbf{B}^{T}$, let us assume $m\geq{n}$ to ease the illustrations. Since we have $\text{rank}(\mathbf{A})=\text{rank}(\mathbf{B})=n$, we can add $(m-n)$ columns for $\mathbf{A}$ and $\mathbf{B}$, i.e., $\tilde{\mathbf{A}}=[\mathbf{A},\mathbf{A}_{1}]$ and $\tilde{\mathbf{B}}=[\mathbf{B},\mathbf{B}_{1}]$, where $\tilde{\mathbf{A}}$ and $\tilde{\mathbf{B}}$ are both nonsingular.\par 
Then let us have a look at $\mathbf{A}+c\mathbf{B}$. If $\text{rank}(\mathbf{A}+c\mathbf{B})<n$, it is certain that $\tilde{\mathbf{A}}+c\tilde{\mathbf{B}}$ is singular, which leads to
\begin{equation}
	\label{det_forlowrank}
	\det(\tilde{\mathbf{A}}+c\tilde{\mathbf{B}})=a_{0}+a_{1}c+\cdots+a_{m}c^{m}=0.
\end{equation}
On the other hand, we have $a_{0}=\det(\tilde{\mathbf{A}})\ne0$ since $\tilde{\mathbf{A}}$ is nonsingular. As a result, there are at most $m$ distinct values of $c$ to satisfy \eqref{det_forlowrank}. It also indicates that there are at most $m$ distinct values of $c$ to make $\text{rank}(\mathbf{A}+c\mathbf{B})<n$, which completes the proof of \textbf{Lemma \ref{lemma1_lowrank_finite}}.
\section{Proof of Theorem \ref{th2_modu}}
\label{th2_modu_proof}
If $\text{rank}(\mathbf{J}_{q})<L$, \eqref{fullrank_modu_judge1} and \eqref{diversity_rank_moduonly} do not hold for one-hot difference between $\mathbf{d}_{1}$ and $\mathbf{d}_{2}$ at the $q$-th element. So the necessity is trivial. Then we assume $\text{rank}(\mathbf{J}_{q})=L$ for $\forall q=0,1,\cdots,M-1$. Similar to the proof of \textbf{Theorem \ref{th1_spread}}, we first provide an algorithm to construct the rotation angles from $\phi_{0}$ to $\phi_{M-1}$. Then we provide the explanation for the efficiency of random generation. \par 
At first, $\phi_{0}$ can be set as any value within $[0,2\pi)$, which guarantees the full rank in \eqref{diversity_rank_moduonly} for $\mathbf{z}(q)=\mathbf{z}(q)\delta_{q}$. To determine the rotation angles in order, let us assume that $\phi_{0},\phi_{1},\cdots,\phi_{q-1}$ has been selected to satisfy the full rank in \eqref{diversity_rank_moduonly} for $\forall \mathbf{z}\in\mathcal{B}^{M}\backslash\left\{\mathbf{0}\right\}$ and $\mathbf{z}(q^{\prime})=0$ for $q^{\prime}\geq{q}$. The target of adding $\phi_{q}$ can then be set as ensuring the full rank for $\forall \mathbf{z}\in\mathcal{B}^{M}\backslash\left\{\mathbf{0}\right\}$ and $\mathbf{z}[q^{\prime}]=0$ for $q^{\prime}>q$. To achieve this, $\left[\tilde{\mathbf{H}}_{0}\mathbf{z},\tilde{\mathbf{H}}_{1}\mathbf{z},\cdots,\tilde{\mathbf{H}}_{L-1}\mathbf{z}\right]$ can be rewritten as \eqref{fullrank_proof_step1_divide} at the top of this page, where we should select $\phi_{q}$ to make sure $\mathbf{B}_{\mathbf{z},q-1}+e^{j\phi_{q}}\mathbf{C}_{\mathbf{z},q}$ enjoys full rank. If $\mathbf{z}(q)=0$, it can be guaranteed considering the property of $\phi_{0},\phi_{1},\cdots,\phi_{q-1}$. If $\mathbf{z}(q^{\prime})=0$ for $\forall q^{\prime}<q$, full rank can still be satisfied that since $\mathbf{C}_{\mathbf{z},q}=\mathbf{z}(q)\mathbf{J}_{q}$. Therefore, we can focus on the scenarios when $\mathbf{z}(q)\ne0$ and $\left[\mathbf{z}(0),\mathbf{z}(1),\cdots,\mathbf{z}(q-1)\right]^{T}\ne\mathbf{0}$. In these cases, we have $\text{rank}\left(\mathbf{B}_{\mathbf{z},q-1}\right)=\text{rank}\left(\mathbf{C}_{\mathbf{z},q}\right)=L$. As a result, there are at most $M(|\mathcal{B}|-1)(|\mathcal{B}|^{q-1}-1)$ distinct values of $\phi_{q}$ to enable the rank of $\mathbf{B}_{\mathbf{z},q-1}+e^{j\phi_{q}}\mathbf{C}_{\mathbf{z},q}$ to be lower than $L$ for some cases of $\mathbf{z}$ since $\mathbf{z}$ has $(|\mathcal{B}|-1)(|\mathcal{B}|^{q-1}-1)$ possibilities and there are at most $M$ different values of $\phi_{q}$ to reduce the rank for each case due to \textbf{Lemma \ref{lemma1_lowrank_finite}}. Since $\phi_{q}$ can be selected from $[0,2\pi)$, there are always infinite choices to determine $\phi_{q}$. Therefore, after deciding the value of $\phi_{M-1}$, we can deduce $\left[\tilde{\mathbf{H}}_{0}\mathbf{z},\tilde{\mathbf{H}}_{1}\mathbf{z},\cdots,\tilde{\mathbf{H}}_{L-1}\mathbf{z}\right]$ enjoys full rank, which guarantees the maximum multipath diversity.\par 
When deciding $\phi_{q}$, we just remove at most $M(|\mathcal{B}|-1)(|\mathcal{B}|^{q-1}-1)$ infeasible values from $[0,2\pi)$, which demonstrates that there are infinite choices for $\phi_{0},\phi_{1},\cdots,\phi_{M-1}$. It also indicates that if $\forall \phi_{q}$ is generated as a continuous random variable within $[0,2\pi)$, e.g., the uniform distribution, the probability of achieving full diversity is $1$, which completes the proof of \textbf{Theorem \ref{th2_modu}}.
\bibliographystyle{IEEEtran}
\bibliography{ref-sum}

@ARTICLE{ref_ODDM, author={Lin, Hai and Yuan, Jinhong}, journal={IEEE Trans. Wireless Commun.}, title={{Orthogonal Delay-Doppler Division Multiplexing Modulation}}, year={2022}, volume={21}, number={12}, pages={11024-11037}, doi={10.1109/TWC.2022.3188776}}

@ARTICLE{ref_channel, author={Bello, P.}, journal={IEEE Trans. Commun. Syst.}, title={{Characterization of Randomly Time-Variant Linear Channels}}, year={1963}, volume={11}, number={4}, pages={360-393}, doi={10.1109/TCOM.1963.1088793}}

@book{ref_LTVchannel_book, place={Cambridge}, title={{Fundamentals of Wireless Communication}}, DOI={10.1017/CBO9780511807213}, publisher={Cambridge University Press}, author={Tse, David and Viswanath, Pramod}, year={2005}}

@ARTICLE{ref_AFDM, author={Bemani, Ali and Ksairi, Nassar and Kountouris, Marios}, journal={IEEE Trans. Wireless Commun.}, title={{Affine Frequency Division Multiplexing for Next Generation Wireless Communications}}, year={2023}, volume={22}, number={11}, pages={8214-8229}, doi={10.1109/TWC.2023.3260906}}

@ARTICLE{OTFS_diversity_TWC, author={Surabhi, G. D. and Augustine, Rose Mary and Chockalingam, A.}, journal={IEEE Trans. Wireless Commun.}, title={{On the Diversity of Uncoded OTFS Modulation in Doubly-Dispersive Channels}}, year={2019}, volume={18}, number={6}, pages={3049-3063}}

@ARTICLE{OTFS_diversity_WCL, author={Raviteja, P. and Hong, Yi and Viterbo, Emanuele and Biglieri, Ezio}, journal={IEEE Wireless Commun. Lett.}, title={{Effective Diversity of OTFS Modulation}}, year={2020}, volume={9}, number={2}, pages={249-253}, doi={10.1109/LWC.2019.2951758}}

@ARTICLE{VandermondeLCP_OFDM_delayOnly, author={Zhiqiang Liu and Yan Xin and Giannakis, G.B.}, journal={IEEE Trans. Commun.}, title={{Linear constellation precoding for OFDM with maximum multipath diversity and coding gains}}, year={2003}, volume={51}, number={3}, pages={416-427}, doi={10.1109/TCOMM.2003.809791}}

@ARTICLE{phase_rota_DFTsOFDM, author={Goeckel, D.L. and Ananthaswamy, G.}, journal={IEEE Trans. Commun.}, title={{On the design of multidimensional signal sets for OFDM systems}}, year={2002}, volume={50}, number={3}, pages={442-452}, doi={10.1109/26.990906}}

@ARTICLE{FrequencyNonzero_TIT, author={{Z. Wang and G. Giannakis}}, journal={IEEE Trans. Inf. Theory}, title={{Complex-field coding for OFDM over fading wireless channels}}, year={2003}, volume={49}, number={3}, pages={707-720}, doi={10.1109/TIT.2002.808101}}

@INPROCEEDINGS{FrequencyNonzero_first_SPAWC, author={Zhengdao Wang and Giannakis, G.B.}, booktitle={Proc. IEEE 3rd Int. Workshop Signal Process. Adv. Wireless Commun. (SPAWC)}, title={{Linearly precoded or coded OFDM against wireless channel fades?}}, year={2001}, doi={10.1109/SPAWC.2001.923899}}

@ARTICLE{TFdouble_Vandermonde, author={Xiaoli Ma and Giannakis, G.B.}, journal={IEEE Trans. Inf. Theory}, title={{Maximum-diversity transmissions over doubly selective wireless channels}}, year={2003}, volume={49}, number={7}, pages={1832-1840}, doi={10.1109/TIT.2003.813485}}

@INPROCEEDINGS{noICI_TVOFDM, author={Zemen, Thomas and Hofer, Markus and Löschenbrand, David and Pacher, Christoph}, booktitle={Proc. IEEE 29th Annu. Int. Symp. Pers., Indoor Mobile Radio Commun. (PIMRC)}, title={{Iterative Detection for Orthogonal Precoding in Doubly Selective Channels}}, year={2018}, doi={10.1109/PIMRC.2018.8580716}}

@ARTICLE{group_OFDM, author={Prasad, Narayan and Venturino, Luca and Wang, Xiaodong}, journal={IEEE Trans. Inf. Theory}, title={{Diversity-Multiplexing Tradeoff Analysis for OFDM Systems With Subcarrier Grouping, Linear Precoding, and Linear Detection}}, year={2010}, volume={56}, number={12}, pages={6078-6096}, doi={10.1109/TIT.2010.2079552}}

@ARTICLE{CP_SC_diversity1, author={Zhang, Wei}, journal={IEEE Trans. Inf. Theory}, title={{Comments on ``Maximum diversity in single-carrier frequency-domain Equalization''}}, year={2006}, volume={52}, number={3}, pages={1275-1277}, doi={10.1109/TIT.2005.864446}}

@ARTICLE{diversity_multiplex_tradeoff_ISI, author={Grokop, Leonard H. and Tse, David N. C.}, journal={IEEE Trans. Inf. Theory}, title={{Diversity–Multiplexing Tradeoff in ISI Channels}}, year={2009}, volume={55}, number={1}, pages={109-135}, doi={10.1109/TIT.2008.2008120}}

@ARTICLE{rotation_VOFDM, author={Han, Chenggao and Hashimoto, Takeshi and Suehiro, Naoki}, journal={IEEE Trans. Commun.}, title={{Constellation-rotated vector OFDM and its performance analysis over rayleigh fading channels}}, year={2010}, volume={58}, number={3}, pages={828-838}, doi={10.1109/TCOMM.2010.03.080291}}

@ARTICLE{TSP_multipath_VOFDM, author={Xiao, Zhenyu and Xia, Xiang-Gen and Bai, Lin},journal={IEEE Trans. Signal Process.}, title={{Achieving Antenna and Multipath Diversities in GLRT-Based Burst Packet Detection}}, year={2015}, volume={63}, number={7}, pages={1832-1845}, doi={10.1109/TSP.2015.2401538}}

@ARTICLE{TSP_multipath_OFDM, author={Tepedelenlioglu, C. and Challagulla, R.}, journal={IEEE Trans. Signal Process.}, title={{Low-complexity multipath diversity through fractional sampling in OFDM}}, year={2004}, volume={52}, number={11}, pages={3104-3116}, doi={10.1109/TSP.2004.836452}}

@ARTICLE{ZP_fullDiversity, author={Rajashekar, Rakshith and Hari, K.V.S. and Hanzo, L.}, journal={IEEE Trans. Commun.}, title={{Spatial Modulation Aided Zero-Padded Single Carrier Transmission for Dispersive Channels}}, year={2013}, volume={61}, number={6}, pages={2318-2329}, doi={10.1109/TCOMM.2013.043013.130011}}

@ARTICLE{noISI_TIT, author={Ke Liu and Kadous, T. and Sayeed, A.M.}, journal={IEEE Trans. Inf. Theory}, title={{Orthogonal time-frequency signaling over doubly dispersive channels}}, year={2004}, volume={50}, number={11}, pages={2583-2603}, doi={10.1109/TIT.2004.836931}}

@ARTICLE{NGcommunication_multipath, author={Miao, Haiyang and Zhang, Jianhua and Tang, Pan and Tian, Lei and Zhao, Xinyu and Guo, Bolun and Liu, Guangyi}, journal={IEEE J. Sel. Areas Commun.}, title={{Sub-6 GHz to mmWave for 5G-Advanced and Beyond: Channel Measurements, Characteristics and Impact on System Performance}}, year={2023}, volume={41}, number={6}, pages={1945-1960}, doi={10.1109/JSAC.2023.3274175}}

@ARTICLE{shuangyang_OTFS, author={Li, Shuangyang and Yuan, Jinhong and Yuan, Weijie and Wei, Zhiqiang and Bai, Baoming and Ng, Derrick Wing Kwan}, journal={IEEE Trans. Wireless Commun.}, title={{Performance Analysis of Coded OTFS Systems Over High-Mobility Channels}}, year={2021}, volume={20}, number={9}, pages={6033-6048}, doi={10.1109/TWC.2021.3071493}}

@ARTICLE{OTFS_diversity_TVT, author={Bora, Amit Sravan and Phan, Khoa T. and Hong, Yi}, journal={IEEE Trans. Veh. Technol.}, title={{Diversity Analysis of OTFS Over Block Time-Varying Channels}}, year={2024}, volume={73}, number={9}, pages={14062-14067}, doi={10.1109/TVT.2024.3396490}}

@ARTICLE{OTFS_diversity_LP_WCL, author={Ge, Yao and Meng, Lingsheng and G., David González and Wen, Miaowen and Guan, Yong Liang and Fan, Pingzhi}, journal={IEEE Wireless Commun. Lett.}, title={{Linear Precoding Design for OTFS Systems in Time/Frequency Selective Fading Channels}}, year={2025}, volume={14}, number={3}, pages={816-820}, doi={10.1109/LWC.2024.3524459}}

@ARTICLE{AFDM_diversity_JSAC, author={Tao, Yiwei and Wen, Miaowen and Ge, Yao and Li, Jun and Basar, Ertugrul and Al-Dhahir, Naofal}, journal={IEEE J. Sel. Areas Commun.}, title={{Affine Frequency Division Multiplexing With Index Modulation: Full Diversity Condition, Performance Analysis, and Low-Complexity Detection}}, year={2025}, volume={43}, number={4}, pages={1041-1055}, doi={10.1109/JSAC.2025.3531561}}

@ARTICLE{OTFS_STSK_diversity, author={Sui, Zeping and Zhang, Hongming and Sun, Sumei and Yang, Lie-Liang and Hanzo, Lajos}, journal={IEEE Trans. Commun.}, title={{Space-Time Shift Keying Aided OTFS Modulation for Orthogonal Multiple Access}}, year={2023}, volume={71}, number={12}, pages={7393-7408}, doi={10.1109/TCOMM.2023.3314861}}

@ARTICLE{PAPR_rota_ref, author={Kim, Jubum and Yun, Yeo Hun and Kim, Chanhong and Cho, Joon Ho}, journal={IEEE Trans. Veh. Technol.}, title={{Minimization of PAPR for DFT-Spread OFDM With BPSK Symbols}}, year={2018}, volume={67}, number={12}, pages={11746-11758}}

@ARTICLE{ref_OTSM, author={Thaj, Tharaj and Viterbo, Emanuele and Hong, Yi}, journal={IEEE Trans. Wireless Commun.}, title={{Orthogonal Time Sequency Multiplexing Modulation: Analysis and Low-Complexity Receiver Design}}, year={2021}, volume={20}, number={12}, pages={7842-7855}}

@ARTICLE{ref_PAPR_oversampling, author={Chin-Liang Wang and Yuan Ouyang}, journal={IEEE Trans. Signal Process.}, title={{Low-complexity selected mapping schemes for peak-to-average power ratio reduction in OFDM systems}}, year={2005}, volume={53}, number={12}, pages={4652-4660}}

@ARTICLE{ref_AMP, author={Yuan, Zhengdao and Liu, Fei and Yuan, Weijie and Guo, Qinghua and Wang, Zhongyong and Yuan, Jinhong}, journal={IEEE Trans. Wireless Commun.}, title={{Iterative Detection for Orthogonal Time Frequency Space Modulation With Unitary Approximate Message Passing}}, year={2022},volume={21}, number={2}, pages={714-725}}

@standard{ref_TDL, title={{NR; User Equipment ({UE}) radio transmission and reception; Part 4: Performance requirements}}, howpublished = {3GPP Specification}, month=sep,year=2024,note={3GPP TS 38.101-4 V18.5.0}}

@ARTICLE{ref_LDPC, author={Richardson, Tom and Kudekar, Shrinivas}, journal={IEEE Commun. Mag.}, title={{Design of Low-Density Parity Check Codes for 5G New Radio}}, year={2018}, volume={56}, number={3}, pages={28-34}, doi={10.1109/MCOM.2018.1700839}}

\vfill
	
\end{document}